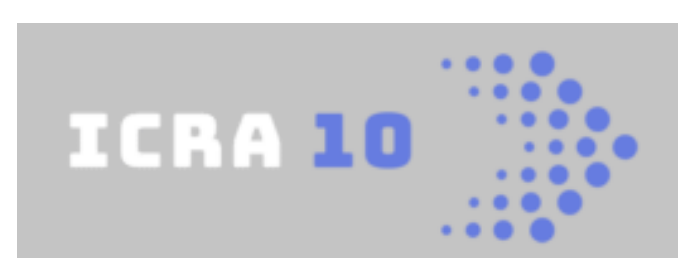

*Study*

# Hidden Markov Model-Based Remaining Useful Life Estimation of Rolling Bearings Using Vibration Signals: A feasibility study

**I. Ksoulos, D.M. Bourdalos, J.S. Sakellariou, S. Malefaki**

**Abstract**: This feasibility study presents a Hidden Markov Model (HMM)-based methodology for the Remaining Useful Life (RUL) estimation of rolling element bearings when the training vibration signals differ significantly from those of the target bearing. An AutoRegressive (AR) model is first employed to capture the machinery dynamics using vibration acceleration measurements from an initial operating period where all components including the considered bearing are still under healthy condition. The AR model is subsequently used to filter a newly acquired vibration signal from the faulty machinery, and typical Envelope Analysis is performed on the residual signal to identify fault-related repetition frequencies. The amplitudes of these frequencies, combined with statistical features of the AR residuals are fused through Principal Component Analysis to construct a sensitive to bearing degradation Condition Indicator (CI). Based on training data from the healthy state, a threshold is established, and a fault is declared when the CI exceeds this threshold. Once a fault is detected, its progression is modelled as a sequence of consecutive, distinct states, which are identified using K-Means clustering. A left-right HMM with continuous observation densities is estimated to represent the fault progression and thus predict the RUL. The HMM-based RUL estimation methodology is trained using vibration measurements from a limited number of two run-to-failure experiments with two nominally identical bearings, while RUL estimation performance is assessed on a third bearing of the same type. Despite the significant differences between the vibration data used for training and the life cycles of the training bearings with respect to the target bearing, the RUL estimation results indicate an adequate but conservative performance of the postulated HMM-based methodology.

**Keywords:** Remaining Useful Life Estimation; Hidden Markov Model; Envelope Analysis; Vibration Signals; Condition Indicator; Rolling Bearing; Rotating Machinery; AR modelling

## 1. Introduction

Rolling bearings are critical components in rotating machinery, widely used in various applications such as electric motors, turbines and manufacturing equipment. However, they are typically subjected to diverse operating conditions, such as overloading, misalignment and inadequate lubrication (Sharma et al., 2015), which makes them particularly prone to incipient faults and eventual failures. Therefore, monitoring their health condition and accurately estimating their Remaining Useful Life (RUL) are of great importance for reliable operation and effective maintenance planning. Accurate RUL estimation allows for maximally efficient component utilization, reduces the operating costs and supports timely replacement decisions, while mitigates the risk of unexpected faults and collateral failure in neighbouring components (Yin et al., 2025).

Vibration-based RUL estimation methods can be broadly categorized into three classes: (a) Physics model-based methods, (b) data driven model-based statistical methods and (c) Neural Network-based methods.

Physics model-based methods describe the degradation process of rotating machinery by building mathematical models based on physics laws and failure mechanisms. The parameters of such models correlate mainly to material properties and rely heavily on assumptions and empirical knowledge. They are generally defined by using specific experiments, finite element analysis or other suitable techniques (Lei et al., 2018). Model driven methods based on physics laws such as Paris-Erdogan law (Yin et al., 2025), Lundberg-Plamgren-based life model (Zhou et al., 2022) and fatigue theoretical design method (Sen et al., 2017) have been used to describe the degradation process by physically modelling the damage been applied mechanisms and have for RUL estimation in rolling bearings. Such models can provide accurate RUL estimation if the physics model is developed with complete understanding of the failure mechanisms (Lei et al., 2018).

**I. Ksoulos, email: *up1072412@upatras.gr* ; D.M. Bourdalos, email: dimitrios.bourdalos@ac.upatras.gr**
**J.S. Sakellariou, email: sakj@mech.upatras.gr ; S. Malefaki, email: smalefaki@upatras.gr**
**Department of Mechanical Engineering & Aeronautics, University of Patras, 26500 Patras, Greece**

However, they rely on a lot of assumptions and expert knowledge and thus offer low accuracy for complex mechanical systems due to model simplifications and approximations (Yin et al., 2025). More recently, physics models have been utilized in hybrid approaches paired with neural networks (NNs) in an effort to reduce the dependencies on expert knowledge and model parameter estimation accuracy (Yin et al., 2025). In this direction, domain adaptation networks (Cui et al., 2024), physics-informed neural networks (Gong et al., 2025; Zhao et al., 2025) and long short-term memory (LSTM) networks (Lu et al., 2024) have been paired with physics models such as the Weibull failure model to estimate RUL of rolling bearings.

Moreover, pure Neural Network-based approaches have gained widespread attention and research in the field RUL estimation of bearings. Such approaches learn degradation patterns from available data using mainly deep learning approaches. These methods prove efficient in dealing with complex mechanical systems whose degradation processes are difficult to interrelate by physics (Lei et al., 2018). Different types of deep learning networks have been examined recently, such as variational Autoencoders (Yang et al., 2024) and hybrid Transformer-Gated Recurrent Units (Wenping et al., 2025; Qu et al., 2025; Xiaochao et al., 2025). Despite promising RUL estimation performance, these methods depend heavily on large datasets and are predominantly evaluated on the same bearing used for training, limiting their practical application. Moreover, they function as low transparency “black boxes” offering little insight into their decision-making process (Lei et al., 2018). Domain experts remain reluctant to rely on models they cannot fully understand or audit (Saeed, et al., 2025).

On the other hand, statistical methods utilizing data driven models offer an attractive alternative type of methods for RUL estimation in rolling bearings. The key advantage of data driven models is that they offer greater explainability than NNs, rooted in empirical functions and probability theory (Kuzio et al., 2025) and are effective in describing the uncertainty of the degradation process (Lei et al., 2018). Prominent approaches include data-driven models based on probability density functions (pdf) and historical failure data to estimate degradation trends and predict RUL, such as Wiener process models, Markov models, Gamma process models and Kalman filters. More specifically, Wiener process models are extensively used to capture non-monotonic fluctuations in Condition Indicators (CIs) (Shang, et al., 2025), often appearing in two-stage prediction frameworks (Galli et al., 2024) or coupled with deep learning models to robustly describe degradation trends (Qinlguan et al., 2024). In contrast, Markov model approaches treat degradation as a transition between finite states, including Hidden Markov Models (HMM) different types of HMMs have been applied for bearing RUL; specifically, mixture of Gaussians HMM (Tobon-Mejia et al., 2012), multi-branch HMM (Galli, et al., 2024) and duration-dependent HMM (Wang et al., 2014). Furthermore, Hidden Semi-Markov Models (HSMM) are applied to relax the restrictive assumptions of state duration found in standard HMMs (Wu et al., 2021). Alternatively, Gamma process models operate on the assumption of independent increments following a gamma distribution, making them suitable for dynamic RUL prediction even when signals exhibit time-varying characteristics or non-Gaussian noise (Kuzio et al., 2025; Wang et al., 2021). Furthermore, Kalman filters and their variations, such as the Adaptive Kernel or Switching Unscented Kalman Filter, are widely utilized to handle nonlinearities and real-time transition probabilities for RUL estimation in rolling bearings (Chen et al., 2024; Li et al., 2025).

However, a common limitation of the above methods lies in their inability to perform well across different bearings, as discrepancy in degradation patterns between the training set and inspection set is commonly observed. This is a common issue, because the degradation trend of rolling bearings varies, even under the same working conditions (Li et al., 2024). Several attempts have been made to apply trained models to an unseen target bearing, such as an Extended Kalman Filter approach (Singleton et al., 2015) and a mixture of Gaussian HMM approach (Tobon-Mejia et al., 2012). These approaches achieved satisfactory results only towards the final stages of bearing RUL. More recently, this problem has been addressed by domain adaptation learning (Li et al., 2024; Zhu et al., 2020; Ze-Jian et al., 2024), which, however, often requires a small amount of data from the unseen bearing for training.

In the present study, the above challenge is investigated via a statistical Hidden Markov Model-based RUL estimation methodology, without using data from the targeted bearing. Regarding the application of the presented methodology, neither multiple datasets nor domain-specific expert input are readily available. More specifically, the identification of that model relies on vibration measurements from two run-to-failure experiments via a single accelerometer, while the effectiveness of the presented methodology is examined on a third unseen bearing, all experiencing the same fault type under the same operating conditions. The key challenge of this study lies in the limited availability of training data from only two bearings of the same type as the target one, but with significantly different wear patterns observed amongst bearings. Within the methodology, initially, an AutoRegressive (AR) modelling technique is employed to remove healthy dynamics from the signals, keeping the fault related to AR residuals. Afterwards, classic Envelope analysis is implemented to extract a frequency domain feature, while time domain features are also derived from the AR residuals. Principal Component Analysis (PCA) is further employed aiming to construct a proper CI by combining time and frequency features. The lifetime of the bearing is then separated into distinct hidden states (HSs) by a predefined healthy threshold based on initial sampling instances and k-means clustering. A proper HMM is finally trained to estimate the RUL of an unobserved bearing. Standard HMM-based RUL estimation for bearings faces several inherent limitations, including high computational complexity, a heavy reliance on large training datasets (Lei et al., 2018), and difficulties in selecting optimal hidden states (Tobon-Mejia et al., 2012). Furthermore, the strict first-order Markov assumption fails to capture complex, long-term degradation dependencies (Wang et al., 2014). To mitigate this specific issue, a time-dependent RUL estimation methodology is utilized in the framework of this study. Finally, the performance of these models is highly sensitive to the accurate determination of the first predicting time (FPT) (Wang et al., 2014).

## 2. Methodology

The RUL estimation methodology can be divided into five steps, as illustrated in Fig. 1. The first step involves signal preprocessing where Autoregressive (AR) modelling is utilized initially to obtain the model residuals, which are then employed for standard Envelope Analysis (Randall & Antoni, 2011, pp. 506-508). In steps 2 and 3, the CI is selected, and the Health State separation is conducted, respectively. In step 4, the HMM is iteratively trained to model the degradation process. Finally, in step 5, the HMM is employed to estimate RUL. The training of the methodology (step 4) is conducted using vibration signals from multiple run-to-failure experiments on a specific bearing type, which constitutes the minimum data requirement for the application of the presented methodology. The performance assessment of the methodology (step 5) is performed on a new run-to-failure experiment with a target bearing of the same type as the training set.

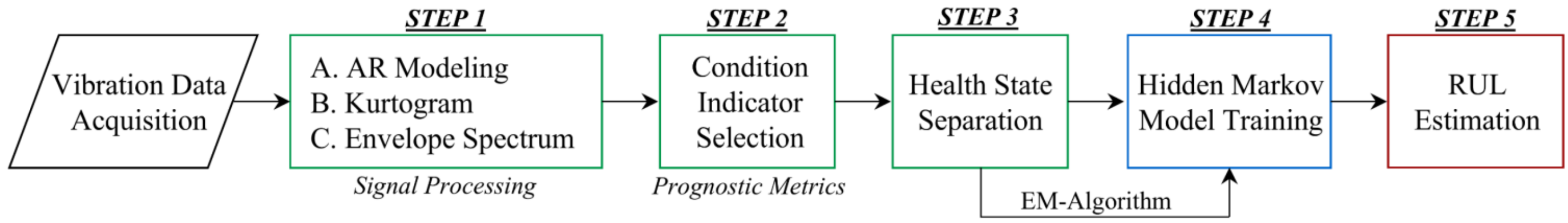

**Fig. 1** Flowchart of the HMM-based RUL estimation methodology

### Step 1: Signal Pre-processing

A. AR Modelling/Residuals: An AR model is first identified to represent the machinery healthy dynamics, using the first available vibration segment of the run-to-failure experiment, which is acquired when all components, including the monitored rolling bearing, are assumed to be fault-free. The identified AR model is then used as a filter to subsequent vibration signals. This filtering enhances bearing fault-characteristic frequencies, making emerging fault signatures more distinct in the model residuals (Randall & Antoni, 2011, pp. 493-494).

B. Kurtogram: Following the AR filtering, the Kurtogram is used to determine an appropriate frequency band for envelope analysis based on spectral kurtosis (SK) (Randall & Antoni, 2011, pp. 501-505). SK quantifies the degree of impulsiveness (non-Gaussian, transient content) as a function of frequency and is therefore well suited to bearing signals where localized defects generate short-duration impacts. The Kurtogram provides a two-dimensional map of SK versus center frequency and analysis bandwidth and the band that maximizes SK is selected to design the band-pass filter for subsequent demodulation (Antoni, 2007).

C. Envelope Spectrum: The AR residuals are then band-pass filtered at the selected frequency bands (determined through Kurtogram). The envelope of the filtered residuals is then obtained through demodulation via the Hilbert transform (Randall & Antoni, 2011, pp. 506-508). Finally, the Fast Fourier Transform (FFT) of the envelope yields the envelope spectrum, which is employed in the subsequent steps of the methodology.

### Step 2: Condition Indicator Selection

Two time-domain features are extracted from the AR residuals and considered as candidate CIs, namely the Peak-to-Peak amplitude (P2P) and the interquartile range (IQR). In addition, a frequency-domain feature is derived from the envelope spectrum of the AR residuals. Specifically, the feature is defined as the maximum magnitude of the envelope spectrum within a predefined frequency bandwidth centered around the expected bearing defect frequency (e.g. Bearing Outer Race Frequency BPFO) and is hereafter referred to as EnvelopeMax. Prior to feature fusion, all features are standardized to zero mean and unit variance to ensure comparability, given their differing scales and physical units. PCA is then applied to selected combinations of the time and frequency domain features, to construct a one-dimensional CI that captures complementary degradation information across domains. Although the number of features is limited, PCA serves as an effective linear transformation for decorrelating the features and emphasizing their shared variance structure, which is directly related to the underlying degradation process. For each feature set, the first principal component – corresponding to the direction of maximum variance – is considered as a CI candidate. Since all the extracted features are driven by the same underlying physical degradation mechanism, they exhibit a consistent increasing trend. As a result, the first principal component reliably captures this dominant behaviour, typically explaining for 85-90% of the total variance. The assumption of linear relationships inherent to PCA is considered reasonable in this context, as the selected features demonstrate strong monotonic relationships during degradation process. Finally, all CI candidates are evaluated using established prognostic metrics namely trendability, prognosability, and robustness (Lei et al., 2018). An overall score is computed as the average of these metrics, and the candidate achieving the highest average score is selected as the final CI.

**Step 3: Health State Separation**

The bearing lifetime is divided into consecutive operating states, based on a multi-stage wear progression model describing the degradation evolution of rolling bearings into distinct phases: Steady-State phase, Defect Initiation, Defect Propagation and Final damage growth (El-Thalji & Erkki, 2014). Accordingly, the employed model in this study defines three states: an initial Healthy state (corresponding to the steady-state phase), a subsequent Degradation state (capturing the physical defect initiation phase), and a final Critical state (capturing the rapid defect propagation and damage growth phases prior to failure) immediately preceding the observable failure. The healthy-state threshold is defined as $\mu + 3\sigma$ (where $\mu$ and $\sigma$ denote the mean and standard deviation of the CI, respectively), corresponding to approximately 99.7% coverage under the Gaussian assumption for data obtained from the initial vibration signals, which are considered representative of healthy operation. When the CI first exceeds this threshold, the bearing is deemed to have entered the faulty regime. All subsequent CI are then partitioned into Degradation and Critical states using K-means clustering. Although K-means is commonly applied in higher-dimensional settings, its application in the present study is restricted to a one-dimensional CI. In this case, K-means effectively reduces to a thresholding mechanism that partitions the CI into distinct degradation regimes, rendering assumptions regarding cluster geometry less critical.

**Step 4: Hidden Markov Model Training**

Based on the above, the fault degradation process can be approximated as a sequence of $N = 3$ consecutive health states (Healthy, Degradation and Critical); therefore, it can be modelled by an HMM in which each hidden state corresponds to a bearing health condition before reaching the final observable failure state. Under the assumption that the degradation process cannot be reversed, a left-right (Bakis type) HMM is utilized (Rabiner, 1989). Such a model only allows transitions to the same state (self-transition) or to the next consecutive (Rabiner, 1989).

Training a left-right HMM requires a set of K multiple observation sequences $\boldsymbol{O} = [O^{(1)}\ O^{(2)}\ ...\ O^{(k)}]$ each with corresponding duration $[T_1\ ...\ T_k]$. Assuming that each observation sequence is independent, the probability of all observation sequences given a single model $\lambda$, $P(\boldsymbol{O}|\lambda)$, is computed by multiplying the individual probabilities $[P_1 .. P_k]$ of each observation sequence (Rabiner, 1989).

Since the CI values are continuous, continuous probability density functions (pdfs) need to be used to model the observation probabilities in each state. For consistent parameter re-estimation, the observation model of each state needs to be defined, in the form of Eq. 2.1, as a finite sum of any log-concave or elliptically symmetric density, in this case Gaussian (Rabiner, 1989) (Juang, 1985).

$$b_j(O^{(l)}) = \sum_{m=1}^{M} c_{jm} f[O^{(l)}, \mu_{jm}, U_{jm}] \tag{2.1}$$

where $O^{(l)}$ is the $l-th$ observed sequence being modelled, $m = [1, M]$ is the number of mixture components and $j = [1, N]$ is the number of states. Hence, $c_{jm}$ is the mixture coefficient for the mth mixture in state j and f is a Gaussian pdf with mean vector $\mu_{jm}$ and covariance matrix $U_{jm}$ in state j. All vectors are assigned with bold lower-case letters and all matrices with bold upper-case letters.

The optimal parameter set for the left-right HMM, taking account multiple observation sequences and continuous observation densities, is iteratively determined by maximizing the probability $P(\boldsymbol{O}|\lambda)$ using the Expectation-Maximization (EM) algorithm, as shown in Eqs. 2.2 - 2.5 (Rabiner, 1989). All estimated parameters are assigned with a hat symbol.

$$\hat{a}_{ij} = \frac{\sum_{k=1}^{K} \frac{1}{P_k} \sum_{t=1}^{T_k-1} a_t^k(i) a_{ij} \mathrm{b_i}\left(\mathrm{O_{t+1}^{(k)}}\right) \beta_{t+1}^k(i)}{\sum_{k=1}^{K} \frac{1}{P_k} \sum_{t=1}^{T_k-1} a_t^k(i) \beta_{t+1}^k(i)} \tag{2.2}$$

$$\hat{\mathrm{c}}_{\mathrm{im}} = \frac{\sum_{\mathrm{k=1}}^{\mathrm{K}} \frac{1}{P_k} \sum_{\mathrm{t=1}}^{T_k} \gamma_{\mathrm{t}}(\mathrm{i,m})}{\sum_{\mathrm{k=1}}^{\mathrm{K}} \frac{1}{P_k} \sum_{\mathrm{t=1}}^{T_k} \sum_{\mathrm{m=1}}^{\mathrm{M}} \gamma_{\mathrm{t}}(\mathrm{i,m})} \tag{2.3}$$

$$\hat{\mu}_{\mathrm{im}} = \frac{\sum_{\mathrm{k=1}}^{\mathrm{K}} \frac{1}{P_k} \sum_{\mathrm{t=1}}^{T_k} \gamma_{\mathrm{t}}(\mathrm{i,m})\, \mathrm{O_t^{(k)}}}{\sum_{\mathrm{k=1}}^{\mathrm{K}} \frac{1}{P_k} \sum_{\mathrm{t=1}}^{T_k} \gamma_{\mathrm{t}}(\mathrm{i,m})} \tag{2.4}$$

$$\hat{\mathrm{U}}_{\mathrm{im}} = \frac{\sum_{\mathrm{k=1}}^{\mathrm{K}} \frac{1}{P_k} \sum_{\mathrm{t=1}}^{T_k} \gamma_{\mathrm{t}}(\mathrm{i,m})\, (\mathrm{O_t^{(k)}} - \mu_{\mathrm{im}})(\mathrm{O_t^{(k)}} - \mu_{\mathrm{im}})'}{\sum_{\mathrm{k=1}}^{\mathrm{K}} \frac{1}{P_k} \sum_{\mathrm{t=1}}^{T_k} \gamma_{\mathrm{t}}(\mathrm{i,m})} \tag{2.5}$$

where $a_t^k(i)$ corresponds to the forward variable and $\beta_t^k(i)$ to the backward variable, considering the $kth$ observation sequence, as computed from the classic forward-backward algorithm. The term $a_{ij}$ corresponds to the transition probability from state $S_i$ to state $S_j$ and $\gamma_{\mathrm{t}}(\mathrm{i,m})$ can be computed by Eq. 2.6 (Rabiner, 1989) and refers to the probability of being in state $S_i$ at time $t$ with the $mth$ mixture component taking account observation $\mathrm{O_t^{(k)}}$, representing the observation at time $\mathrm{t}$ of the $kth$ observation sequence.

$$\gamma_{\mathrm{t}}(\mathrm{i,m}) = \left[\frac{a_t^k(i)\beta_t^k(i)}{\sum_{\mathrm{i=1}}^{\mathrm{N}} a_t^k(i)\beta_t^k(i)}\right]\left[\frac{\mathrm{c_{im}} \mathrm{f}\left[\mathrm{O_t^{(k)}}, \mu_{\mathrm{im}} \mathrm{U_{im}}\right]}{\sum_{\mathrm{m=1}}^{\mathrm{M}} \mathrm{c_{im}} \mathrm{f}\left[\mathrm{O_t^{(k)}}, \mu_{\mathrm{im}} \mathrm{U_{im}}\right]}\right] \tag{2.6}$$

**Step 5: Remaining Useful Life Estimation**

For the RUL estimation step, after estimating the most probable current state through the classic Viterbi algorithm (Rabiner, 1989), the next step is predicting the future state transitions until failure is reached. In that sense, two RUL methods are examined: a time-independent and a time-dependent one.

Time independent: Within the HMM framework, the RUL is defined as the number of state transitions, or number of sampling instances required to reach the final failure state $S_{N+1}$ for the first time from the current state $S_i$. By this definition, RUL can be seen as a discrete variable. Since the model is strictly a left-right HMM its probability mass function (pmf) can be calculated following the backward recursion, in Eqs. 2.7a - 2.7b, for $n$ possible remaining number of state transitions (Thanh, Florent, & Christophe, 2014). For the currently estimated state $S_i$:

$$R\hat{U}L_i^t = P(RUL = 1 | q_t = S_i) = a_{iN} \tag{2.7a}$$

$$R\hat{U}L_i^t = P(RUL = n | q_t = S_i) = a_{ii} R\hat{U}L_i^{(n-1)} + a_{i(i+1)} R\hat{U}L_{i+1}^{(n-1)} \tag{2.7b}$$

Time dependent: The performance of the model with a time-dependent approach is also examined. This method also accounts for the sojourn time $\tau$, in terms of the number of sampling instances the system remains in a specific state $S_i$.

Given the model $\lambda$ and the current estimated state sequence provided by the Viterbi algorithm, the probability of remaining in a state for a period of $d$ sampling instances follows a geometric distribution for an HMM (Rabiner, 1989). The RUL prediction pmf, for every time $t$ can be computed based on Eq. 2.8 (Kontogiannis et al., 2025).

$$R\hat{U}L_i^t = P(RUL = d | q_t = S_i, \{q_1 = S_i, \dots, q_\tau = S_i\}) = d_{i,i}\left(D_i(d-\tau) + \sum_{k=i+1}^{N} D_k(d) + f(1,\varepsilon)\right) +$$
$$+ d_{i,i+1}\left(\sum_{k=i+1}^{N} D_k(d) + f(1,\varepsilon)\right) \tag{2.8}$$

In Eq. 2.8 the term $D_i(d)$ represents the pmf for a state $S_i$ having a duration of $d$ sampling instances, which is geometric. Variable $\tau$ corresponds to the number of sampling instances that the system has already been in state $S_i$ as defined by the state sequence that has been estimated from the Viterbi algorithm until time $t$. Hence, the term $D_i(d-\tau)$ corresponds to the pmf for the period the system stays in state $S_i$ given that it has already remained for $\tau$ sampling instances. The term $\sum_{k=i+1}^{N-1} D_k(d)$ represents the convolution of pmfs, for the duration until $d$ sampling instances, of the future states of the system. Term $f(1,\varepsilon)$ corresponds to the final absorbing failure state of the system. The system remains in that state just once; therefore, it is modelled as a normal distribution with unit mean (duration of one sampling instance) and standard deviation $\varepsilon$. The terms $d_{i,i}$ and $d_{i,i+1}$ correspond to the transition probabilities (Kontogiannis et al., 2025).

## 3. Experimental Setup

The experimental dataset used (XJTU-SY bearing datasets) is provided by the Institute of Design Science and Basic Component at Xian Jiaotong University (XJTU), Shaanxi, P.R. China, and Changxing Sumyoung Technology Co., Ltd. (SY), Zhejiang, P.R. China (Wang et al., 2020). The experimental set-up consists of an AC motor, a motor speed controller, a support shaft, two support bearings (heavy duty rolling bearings), and a hydraulic loading system, as displayed in Fig. *2*. The setup is suitably designed for conducting accelerated degradation tests under different operating conditions (e.g., varying radial load and rotational speed). The radial force is generated by the hydraulic loading system and applied to the housing of each bearing, while the rotational speed of the shaft is adjusted and maintained constant throughout the experiment by the speed controller of the AC motor. Two accelerometers are used to acquire the vibration signals of the tested bearing. The accelerometers are of type PCB 352C33 and are mounted at a 90° angle in horizontal and vertical directions, respectively, on the bearing housing (see Fig. *2*). To observe the complete degradation processes of bearings, each of the accelerated tests was performed until the maximum amplitude of the horizontal or vertical vibration signals exceeded $10 \times A_h$, where $A_h$ is the maximum amplitude of the horizontal or vertical signals in the normal operating stage.

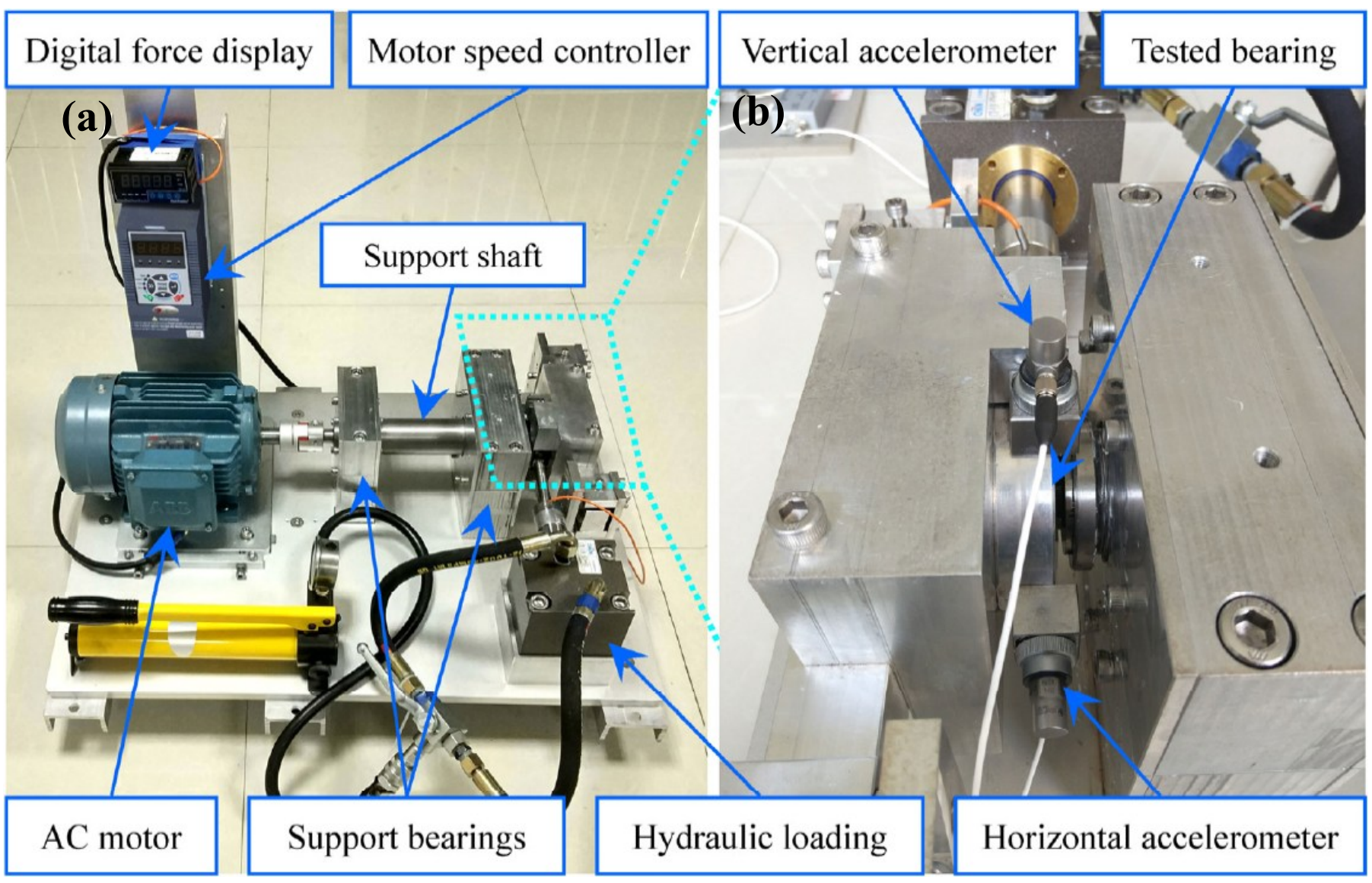

**Fig. 2** Photo of the experimental set-up for the accelerated degradation of rolling element bearings: **(a)** *AC motor, support shaft and bearings & Hydraulic loading system*, **(b)** *Tested bearing & vertical and horizontal accelerometers*

Three run-to-failure experiments are considered in this study. All of them are performed under identical, constant operating conditions, with the rotational speed set to 2.100 rpm and the applied load set to 12 kN. In each experiment, the same bearing type (LDK UER204) is tested, and the three specimens are denoted as Bearing No.1, Bearing No.2, and Bearing No.3. After each experiment, the same failure mode is observed, namely an outer-race defect. The sampling frequency is set to 25.6 kHz. At each sampling instant, a single vibration segment of 32.768 data-points (1.28 sec) is acquired. Consecutive segments are acquired at 60 sec intervals. All the details in the experiments are summarized in Table *1*.

The training of the presented methodology is performed using two run-to-failure experiments (Bearing No.2 and Bearing No.3), whereas RUL estimation performance is evaluated using the remaining experiment (Bearing No.1). This split is adopted because Bearing No. 2 and Bearing No. 3 exhibit comparable end-of-life durations, providing a consistent basis for model training. Only the horizontal accelerometer measurements are used in this study.

**Table 1** Experimental details.

| **Operating Conditions** Speed (rpm) / Load (kN) | **Bearing No.** | **# of Segments** | **Lifetime duration** | **Fault type** | **BPFO**[1] (Hz) |
|---|---|---|---|---|---|
| 2100 / 12 | 1 | 123 | 2 h 3 min | Outer Race | 107.91 |
| | 2 | 161 | 2 h 41 min | | |
| | 3 | 158 | 2 h 38 min | | |

*[1] bearing outer race fault frequency; Sampling frequency:* $f_s$*=25.6 kHz; segment length: 32.768 data-points (1.28 sec); sampling intervals: 60 sec; Training: Bearing No. 2 and No.3; Inspection: Bearing No.1*

## 4. Performance assessment

### Step 1: Signal Pre-Processing

A. AR Modelling/Residuals: Following the procedure described in section 2, an AR model is estimated from the first vibration segment of each run-to-failure experiment. The AR identification follows standard procedures, with the model order selected using the Bayesian Information Criterion (BIC), parameter estimation performed via Ordinary Least Squares (OLS), and validation carried out through typical residual uncorrelatedness (whiteness) testing, which herein is based on the Portmanteau test with $\alpha = 10^{-2}$ and 50 lags (Ljung, 1999, pp. 81-83). Then, the estimated AR models are utilized to filter each subsequent vibration segment, obtaining thus the corresponding AR residuals. The frequency content of the AR residuals is compared with that of the original vibration segments in Fig. *3*. Specifically, for each bearing (different run-to-failure experiment), the spectra are shown over the entire lifetime as a function of frequency and segment No.: subplots (a), (c), and (e) correspond to the original segments of Bearings No. 1–3, respectively, whereas subplots (b), (d), and (f) show the associated residual spectra. In the original segments, strong low-frequency components (approximately 0 – 5.000 Hz) dominate the response and reflect the healthy machinery dynamics. After AR filtering, these components are substantially attenuated, yielding residuals in which the degradation-related spectral content becomes more apparent.

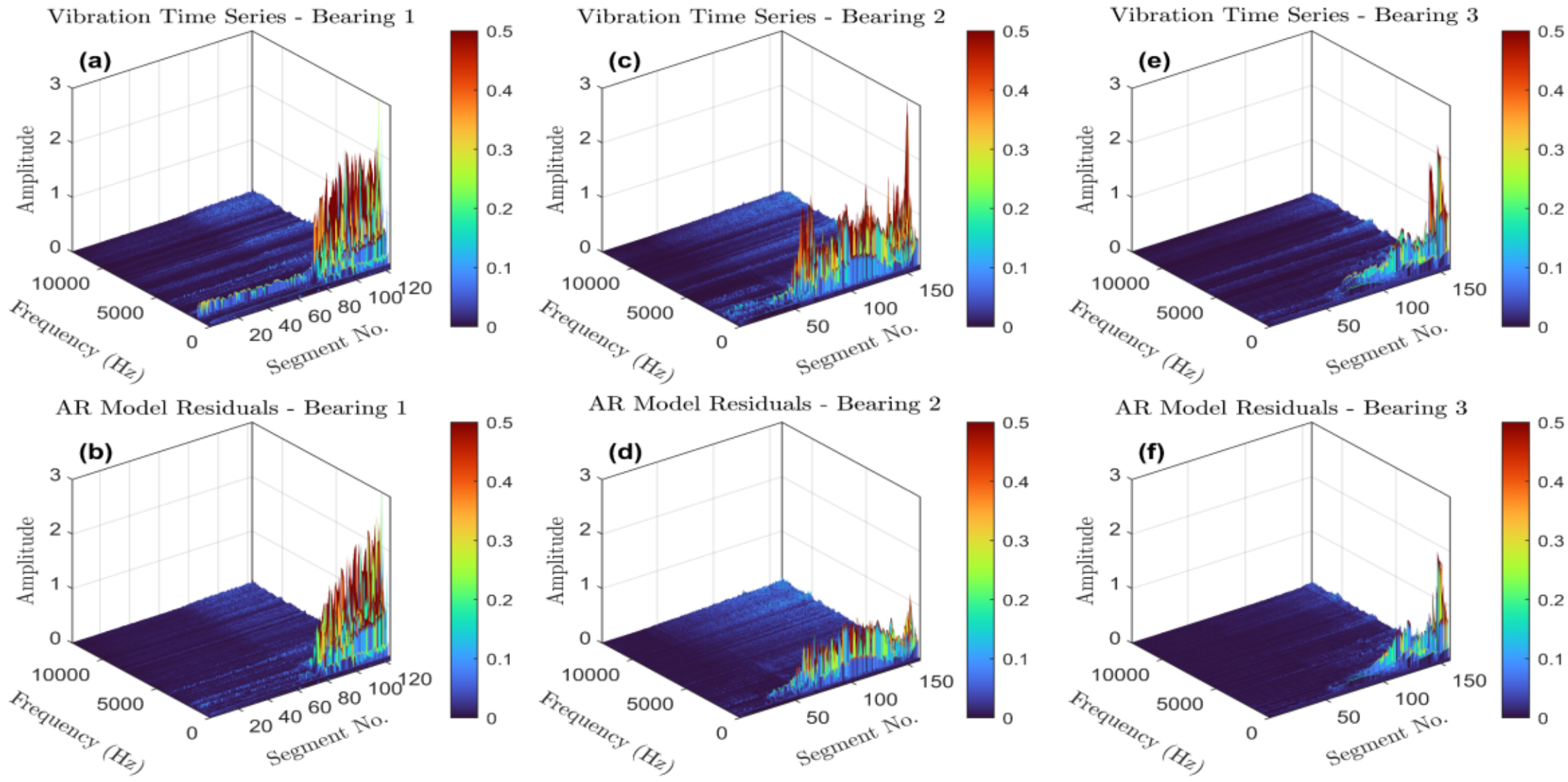

**Fig. 3** FFT amplitude as a function of frequency and segment No. using the vibration time series **(a), (c), (e),** and the corresponding AR model residuals **(b), (d), (f)** across the bearings' lifetime.

B. Kurtogram: Based on the procedure described in section 2, following the AR modelling, the Kurtogram is used to determine an appropriate frequency band for bandpass filtering for subsequent envelope analysis. For each bearing, the Kurtogram is computed from a representative AR-residual segment acquired close to end-of-life, where degradation-related impulsive content is expected to be present. Fig. 4 shows the resulting Kurtograms for (a) Bearing No. 1, (b) Bearing No. 2, and (c) Bearing No. 3. The selected filtering bandwidths (center frequency and bandwidth) are reported in the same figure.

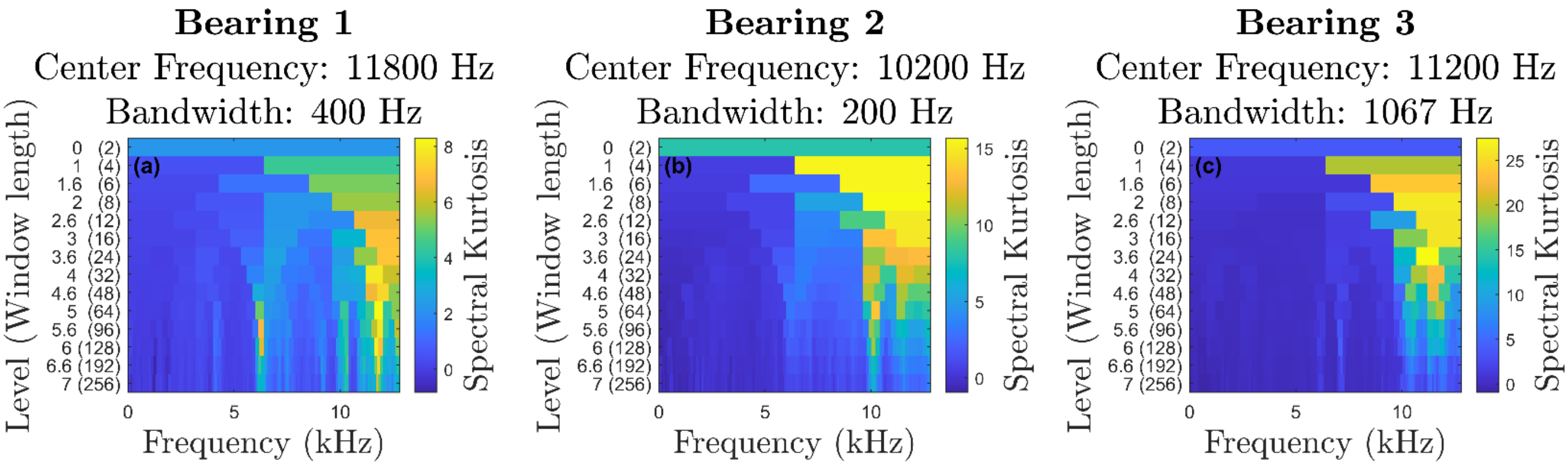

**Fig. 4** Kurtograms computed from one representative AR-residual segment for each run-to-failure experiment: **(a)** Bearing No. 1, **(b)** Bearing No. 2, and **(c)** Bearing No. 3. The color scale indicates spectral kurtosis as a function of center frequency and decomposition level (window length). The selected bandwidth for filtering (center frequency and bandwidth) is reported above each subplot

C. Envelope Spectrum: Finally, the AR residuals are bandpass filtered in the determined via Kurtogram bandwidths and the corresponding envelope spectra are estimated. Fig. 5 shows the envelope spectrum for an indicative segment from each run-to-failure experiment: (a) Bearing No. 1, (b) Bearing No. 2, and (c) Bearing No. 3. In each subplot, the theoretical BPFO is marked with a red asterisk. In all three cases, a clear spectral component is observed at BPFO, confirming that the selected bandwidths and the envelope analysis successfully reveal the expected bearing fault signature.

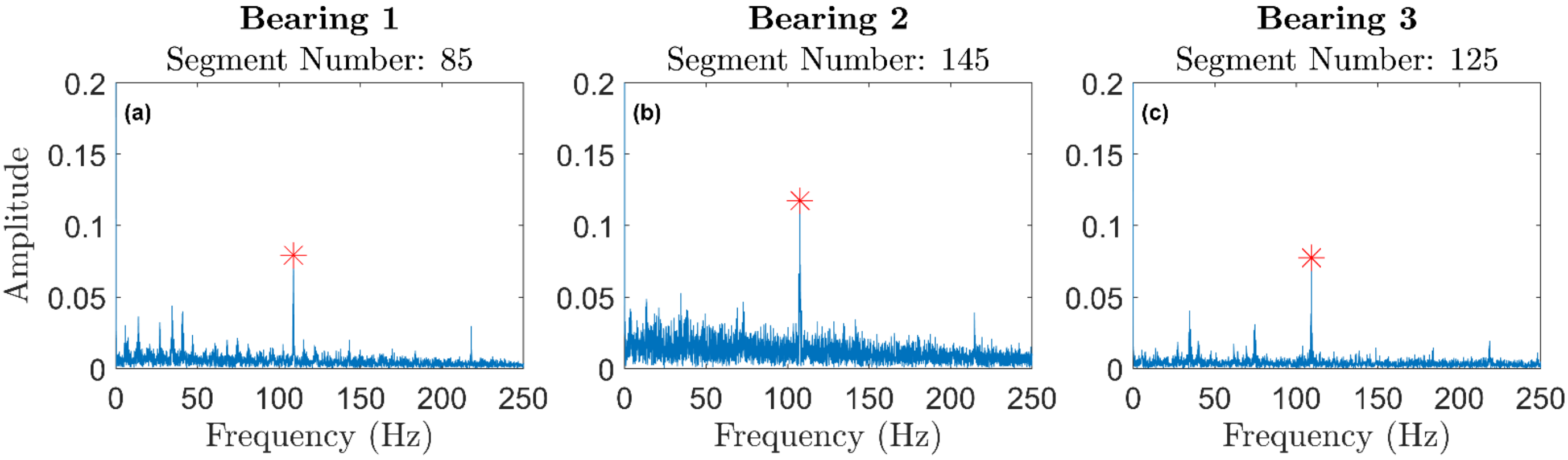

**Fig. 5** Envelope spectra of the bandpass filtered AR residuals for one representative segment from each run-to-failure experiment: **(a)** Bearing No. 1 (Segment 85), **(b)** Bearing No. 2 (Segment 145), and **(c)** Bearing No. 3 (Segment 125). The *red asterisk* denotes the theoretical BPFO

**Step 2: Condition Indicator Selection**

The subsequent step of the presented methodology focuses on the selection of an appropriate CI. In accordance with the procedure outlined in step 2 of Section 2, three features are extracted from the AR residuals as CI candidates, namely the P2P, the IQR, and the EnvelopeMax, along with PCA-derived combinations of these features. Prior to PCA application, all training features are standardized to ensure comparability across variables with heterogeneous scales and units. Specifically, a z-score normalization is employed. During the testing phase, where data are treated as unseen, the same normalization parameters obtained from the training set are applied, thereby ensuring consistency and preventing data leakage. The extracted features from the training bearings (Bearing No. 2 and Bearing No. 3) exhibit strong linear interdependencies, with Pearson correlation coefficients exceeding 0.749. This level of correlation substantiates the suitability of PCA as a linear feature fusion technique, enabling the extraction of latent variables that capture the shared variance associated with the underlying degradation process. For each CI candidate – either an original feature or PCA-derived component - standard prognostic metrics are evaluated. An aggregate performance score is subsequently computed as the mean value of these metrics and is used to identify the most representative and robust CI for the degradation modelling framework. Fig. 6 (a) summarizes the resulting average scores. As can be seen the highest score is achieved by the first principal component obtained from the combined EnvelopeMax and IQR feature set. This component is therefore selected as the final CI. The selected CI trajectories for the three run-to-failure experiments are shown in Fig. 6 (b), with each one indicated by a different colour. As can be seen, the CI does not follow the same pattern across the different experiments even though the same bearing type is tested under identical operating conditions. This variability highlights the highly challenging problem, where the model is to be trained on two run-to-failure cases and its RUL estimation performance is assessed on the third, unseen case.

**Fig. 6** CI Selection: **(a)** Average prognostic metric score (trendability, prognosability, and robustness) for each candidate feature and PCA-based combination. **(b)** Trajectory of the selected CI over the bearing lifetime for the three run-to-failure experiments (Bearings No. 1–3)

**Step 3: Health State Separation**

The third step of the methodology involves the health state separation. As previously mentioned, the 2nd and 3rd run-to-failure experiments involving bearing No.2 and No.3 are used for the training of the presented methodology. Based on the procedure described in Section 2, the Healthy state threshold is defined for each experiment as $\mu + 3\sigma$, where the $\mu$ and $\sigma$ of the CI are computed from the first 20 vibration segments. The selection of the first 20 segments as representative of healthy operation, is based on the experimental design and empirical observations. As the initial measurements correspond to normal operation, they exhibit very low variability, and the use of the $\mu + 3\sigma$ threshold effectively captures the healthy data range. In particular, the maximum deviation of the CI values within the initial 20-segment interval for the training bearings (Bearing No. 2 and Bearing No. 3) is $2.25 \times 10^{-4}$ and $2.57 \times 10^{-4}$, respectively. In addition, each segment corresponds to one minute of operation (60-second sampling interval), so the first 20 segments represent 20 minutes of operation, which empirically ensures a sufficiently stable baseline while avoiding any risk of including early-stage degradation, not expected to occur within this initial period, which are assumed to represent healthy operation. Among the two training experiments the maximum threshold value is selected as global threshold. The Degradation and Critical states are then separated using K-means clustering. This procedure is depicted in Fig. 7 (a) and (b) for the bearing No.2 and No.3, respectively, with blue corresponding to the Healthy State, orange to the Degradation State and red to the Critical State, while the global health state threshold is shown in horizontal black dashed lines.

**Step 4: Hidden Markov Model Training**

For the HMM to parametrically describe the continuous values of the CI for each HS, a Gaussian Mixture Model (GMM) is fitted to the training data corresponding to each state. That model is formulated as in Eq. 2.1. The GMM number of components is determined by the BIC to be two for each state. The resulting pdfs along with the CI values observed for each state are presented in Fig. 8 . The transition matrix of the HMM is initialized randomly in the form of a left-right HMM.

A left–right Hidden Markov Model (HMM) with continuous observation densities is initialized by the parameter set $\boldsymbol{\lambda} = (\boldsymbol{A}, \boldsymbol{c}, \boldsymbol{\mu}, \boldsymbol{U})$, where $\boldsymbol{A}$ represents the state transition probability matrix, and $\boldsymbol{c}$, $\boldsymbol{\mu}$, and $\boldsymbol{U}$ denote the weights, means, and standard deviations of the Gaussian Mixture Model (GMM) used for the continuous observations. For each HS, goodness-of-fit assessment is performed between the empirical training distributions and the corresponding fitted GMMs. Q–Q plots (Fig. 9) are constructed by comparing empirical quantiles with the theoretical quantiles implied by the fitted GMMs, providing a graphical assessment of distributional agreement. In addition, the Kolmogorov–Smirnov test is employed to test whether the empirical data are consistent with the fitted GMM. The null hypothesis is not rejected for any health state, with p-values of 0.7242, 0.8030, 0.7480 respectively, indicating no statistically significant deviation between empirical and model-based distributions. Overall, both graphical diagnostics (Q–Q plots) and formal hypothesis testing consistently support the adequacy of the fitted GMMs for all health states.

**Fig. 7** Health-state separation for the training experiments using the selected CI: **(a)** Bearing No. 2 and **(b)** Bearing No. 3. The dashed line shows the global health state threshold computed from the first 20 segments of each bearing. Samples are labelled Healthy *(blue)*; post-threshold samples are clustered into Degradation *(orange)* and Critical *(red)*

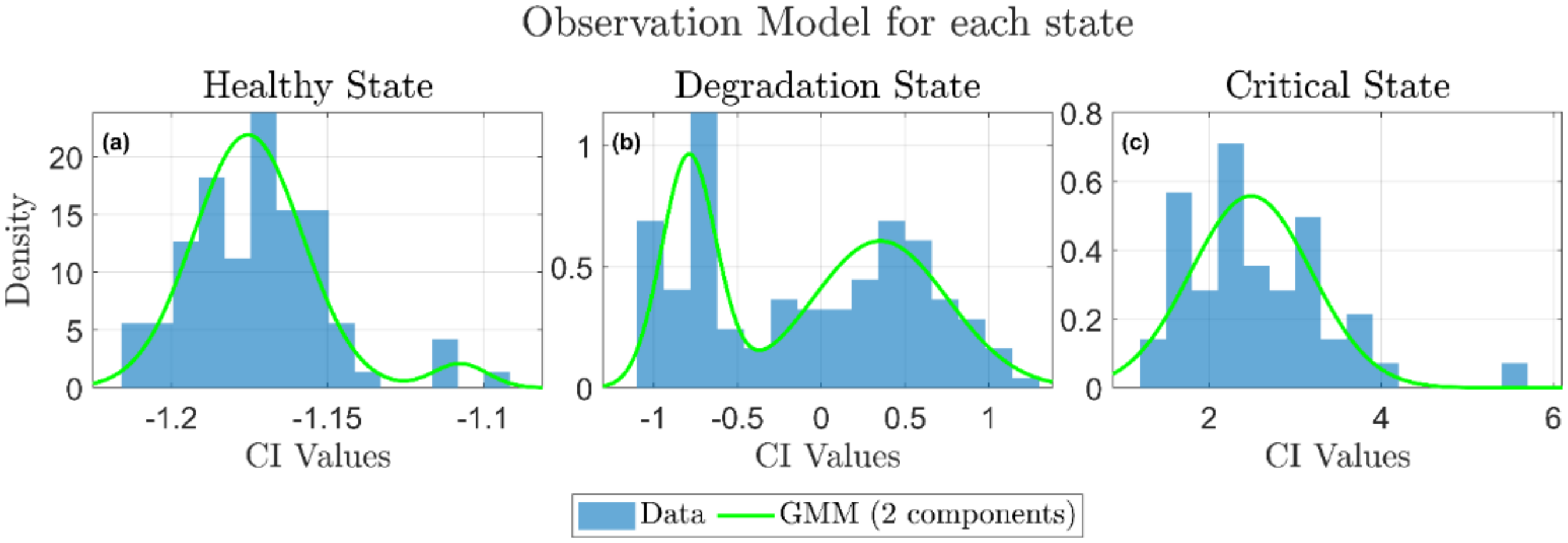

**Fig. 8** Resulted Gaussian Mixture Model (GMM-2 components) pdf (green) estimated using training data from each bearing lifetime state:**(a)** Healthy State, **(b)** Degradation State and **(c)** Critical State

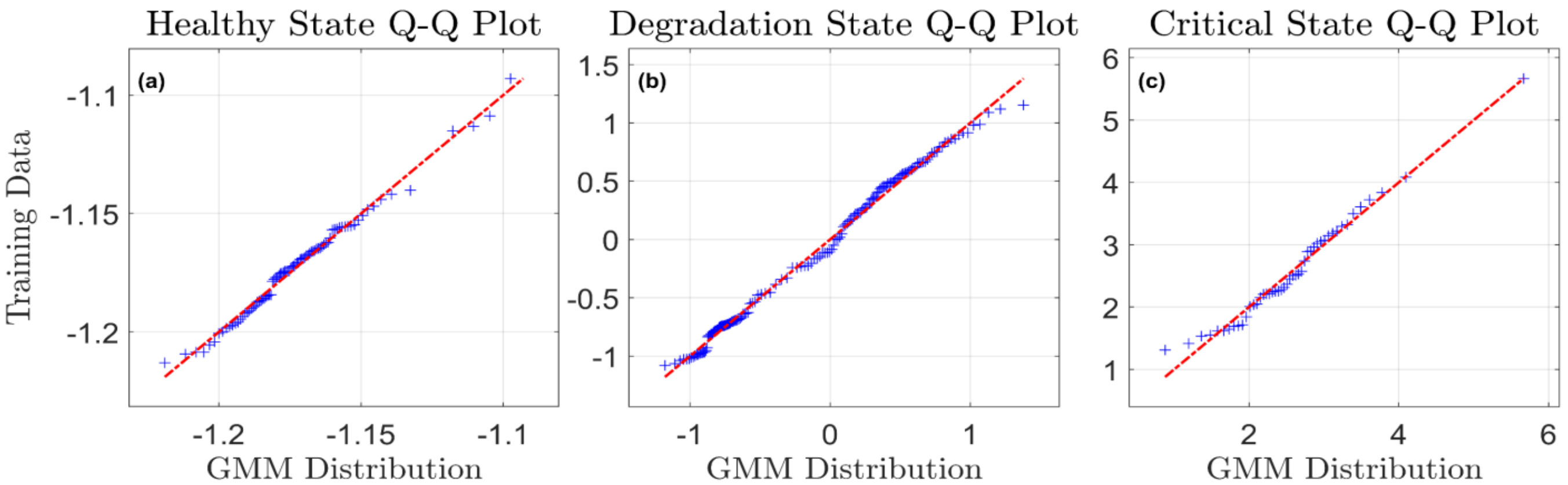

**Fig. 9** GMM goodness-of-fit Q-Q plots for the (a) Healthy, (b) Degradation, and (c) Critical states. The 45-degree reference line (red dash-dotted) is presented to quantify distribution agreement.

The parameters of the HMM are iteratively re-estimated using the Expectation–Maximization (EM) algorithm, according to Eqs. 2.2 - 2.5, as described in Section 2. The observation sequences corresponding to the training bearings are assumed to be statistically independent, as the run-to-failure experiments were conducted separately, with no physical interaction between bearings. At the same time, all bearings are governed by the same underlying physical degradation mechanism, since they are of identical type and operate under the same conditions. The algorithm is initialized with a maximum of 100 iterations, and convergence is assessed based on a tolerance threshold applied to the BIC, which is computed from the likelihood of the observation sequences $\boldsymbol{O}$ given the estimated model $\lambda$.

**Step 5: Remaining Useful Life Estimation**

After the HMM is trained using the run-to-failure experiments of Bearings No. 2 and No. 3, it is applied to Bearing No. 1 for RUL estimation. An online setting is emulated by producing an RUL estimate every 10 minutes (corresponding to 10 segments) over the full lifetime of Bearing No. 1, using data available up to the current segment only. At each step, the most likely current health state is inferred using the Viterbi algorithm. The two RUL estimation approaches described in Section 2 are then evaluated. More specifically, Fig 10. (a) and (b) report the predicted RUL values (blue asterisks) against a linear approximation of the true RUL (red line) for the time-independent and time-dependent methods, respectively. The 95% confidence bounds are shown with blue dotted lines in both subplots and are extracted from the cumulative distribution function (CDF) of the predicted RUL pmf at each prediction step capturing 95% of the probability mass.

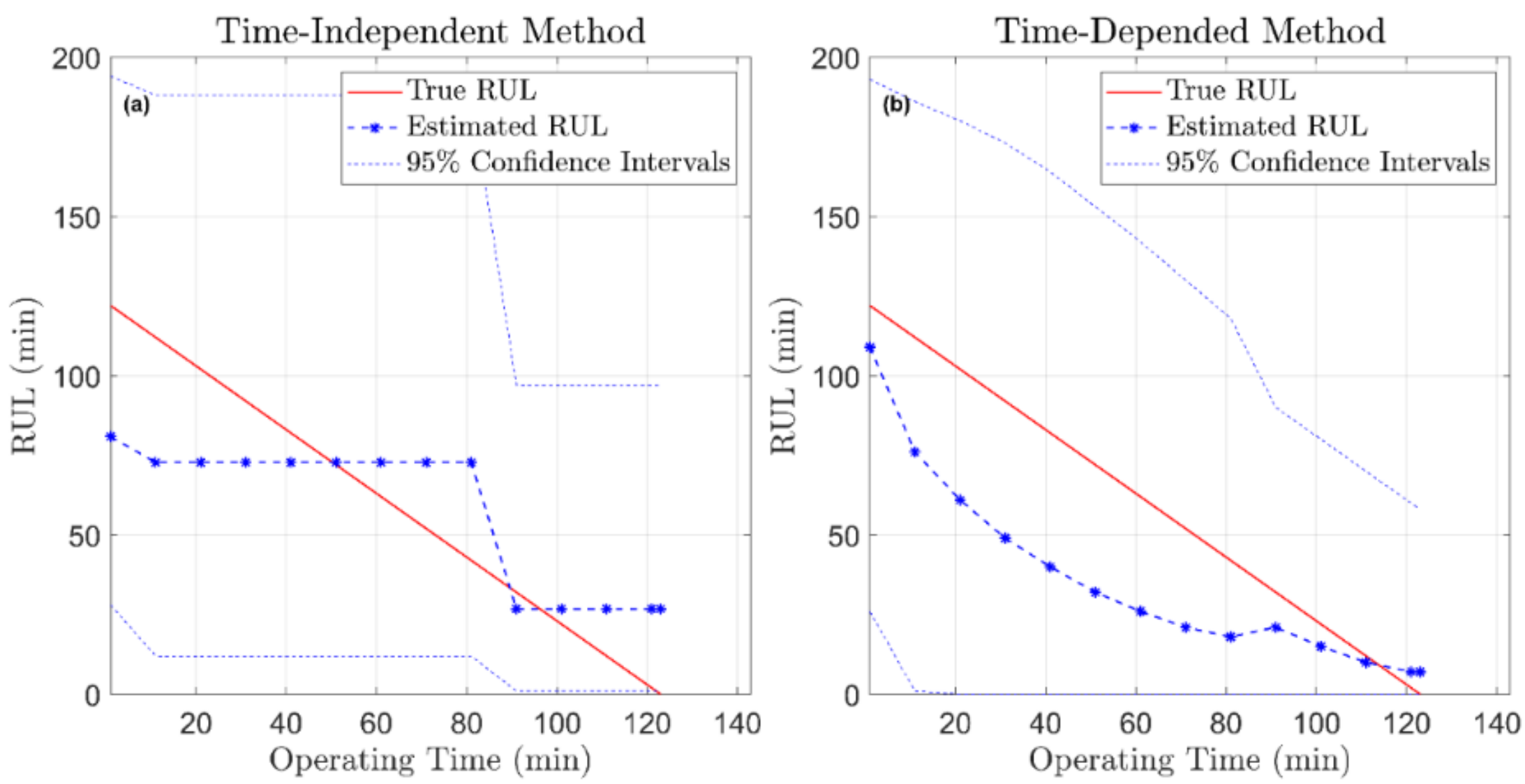

**Fig 10.** RUL estimation results for Bearing No. 1 using the HMM trained on run-to-failure experiments with Bearings No. 2 and No. 3: **(a)** time-independent method and **(b)** time-dependent method. Estimated RUL values are shown with *blue asterisks*, the linear approximation of the true RUL is shown with a *red line*, and the 95% confidence intervals are indicated by *blue dotted lines*

For the time-independent approach, the predicted RUL is computed based on Eqs. 2.7a - 2.7b. As observed in Fig. 10 (a), RUL prediction exhibits a stepwise profile because the estimate depends only on the current Viterbi state; the RUL remains unchanged until a state transition is detected. Consequently, the method provides no gradual RUL updates between consecutive transitions. In contrast, the time-dependent approach, based on Eq. 2.8, uses both the current state and the time spent in that state (sojourn time) and therefore updates the RUL at every prediction step. During the estimated Degradation state (operating time 10–80 min), the predictions are conservative because Bearing No. 1 degrades faster than the training bearings (No. 2 and No. 3) and thus transitions between states earlier than expected by the model. In the Critical state (operating time 80–120 min), the predictions become closer to the true RUL.

## 5. Concluding Remarks

A HMM-based methodology for the RUL estimation of rolling element bearings, using training vibration signals which significantly differ from those of the target bearing, along with its experimental assessment, has been presented in this study. The methodology has been trained using single-axis accelerometer measurements from two run-to-failure experiments on rolling bearings of the same type under constant operating conditions, and assessed on a third, previously unseen target bearing, with the introduction of a new CI to track degradation.

Substantial differences in degradation behaviour are observed between the two bearings used for training, making RUL estimation of the unseen bearing highly challenging. However, without using data from the target bearing for training, the postulated methodology provided adequate, albeit conservative, RUL estimates.

The conservative nature of the predictions indicates that further refinement is required to improve accuracy in future work. The methodology corresponds to TRL 4 (component validation in a laboratory environment), as it has been validated using run-to-failure data from a controlled experimental setup. Progression towards higher TRLs will require further assessment under more complex test configurations and varying operating conditions, alongside continued performance improvements.